\documentclass[]{rsos}

\usepackage{graphicx}
\usepackage{xcolor}
\usepackage{adjustbox}        
\begin{document}
\title{Five decades of US, UK, German and Dutch music charts
show that cultural processes are accelerating}

\author{Lukas Schneider$^{1}$, Claudius Gros$^{1}$}
\address{$^{1}$Institute for Theoretical Physics, Goethe University Frankfurt,
Frankfurt a.\,M., Germany}

\subject{e-science, complexity, systems theory}


\corres{Claudius Gros\\
\email{gros07[at]itp.uni-frankfurt.de}}

\begin{abstract} 

Analyzing the timeline of US, UK, German and Dutch 
music charts, we find that the evolution of album 
lifetimes and of the size of weekly rank changes
provide evidence for an acceleration of cultural 
processes. For most of the past five decades number 
one albums needed more than a month to climb to the 
top, nowadays an album is in contrast top ranked 
either from the start, or not at all. Over the last
three decades, the number of top-listed albums increased as a
consequence from roughly a dozen per year to about 40.

The distribution of album lifetimes evolved during 
the last decades from a log-normal distribution
to a powerlaw, a profound change. Presenting an 
information-theoretical approach to human activities, 
we suggest that the fading relevance of personal 
time horizons may be causing this phenomenon. 
Furthermore we find that sales and airplay based 
charts differ statistically and that the inclusion
of streaming affects chart diversity adversely.

We point out in addition that opinion dynamics may accelerate
not only in cultural domains, as found here, but also
in other settings, in particular in politics, 
where it could have far reaching consequences.
\end{abstract}

\begin{fmtext}    

\section{Introduction}

Music charts constitute a valuable source for the study of
extended timelines of culturally and socially relevant
data. One of the most influential collection of 
music charts, the US-based Billboard charts, has 
been used in this context to examine the evolution 
of popular music and to test theories of
cultural change \cite{mauch2015evolution}. Other approaches
concentrated on the fractional representation of race and 
gender \cite{lafrance2018race,flynn2016objectification}, on
the distribution of blockbusters among superstars
\cite{ordanini2016fewer}, on linguistic and 
\end{fmtext}
\maketitle

\noindent
psychological aspects \cite{nishina2017study,nunes2015power}, 
and on the question whether 
there is a trend towards a converging global popular 
music culture \cite{achterberg2011cultural}.
For the UK charts, a correlation analysis between
musical trends, acoustic features and chart success
has been performed \cite{interiano2018musical}.
On a general level the interplay between significance and
popularity has been investigated for the case of
online music platforms \cite{monechi2017significance}.

An especially interesting aspect of music charts is
that they allow to study if and how time scales that
are potentially relevant for cultural and sociological
developments have changed over the last five decades. 
This is a central theme for the theory of social 
acceleration \cite{rosa2013social}, which presumes 
that social and cultural time scales have seen a 
continuing acceleration \cite{vostal2017slowing}.
The pace of time is also a key determinant for liberal 
democracies \cite{scheuerman2004liberal}, which are based 
on reliable temporal ties between politics and electorate 
\cite{goetz2014question}.

Empirical studies attempting to determine
quantitatively the long-term evolution of political,
social or cultural time scales are rare
\cite{ulferts2013acceleration,rosaPrivatComm}.
Here we point out that music charts allow
to investigate the long-term evolution of
a given cultural time scale. For the US, the 
UK, the German and the Dutch charts we find 
that several core chart characteristics, 
such as the overall chart
diversity, the album lifetime and the entry
position of number one albums, indicate that
the pace of the underlying generative processes
has accelerated substantially over the last decades,
by a factor of two or more, in particular since the
rise of the internet. The evolution of the US
Billboard and the German music charts are very
similar, with the Dutch charts showing a time lag
of roughly a decade. The UK charts are on the other
side more conservative, in the sense that their
statistics changed less dramatically since the early
80s.

For the number one albums we find a complete reversal between
the early decades, from the 60s to the 80s, and the
situation as of today. In the past essential no number one
album would start at the top of a chart. Reaching the
top was instead a tedious climbing process that would
take on the average an entire month, or more. Nowadays
the situation is the opposite. If an album is not
the number one the first week of its listing, it has
only a marginal chance to climb to the top later on. 
We believe that these empirical findings constitute
quantitative evidence that the time scales determining
cultural penetration and opinion formation processes have
shortened substantially, in particular since the early 90s.

Besides averaged quantities, we examine in detail the
distribution of album lifetimes. The probability distribution
that an album is listed overall for a certain number of weeks
has seen a conspicuous evolution over the last 3-4 decades,
with a log-normal distribution changing continuously into
a powerlaw. This evolution can be interpreted as a
self-organizing process unfolding slowly over the course
of several decades. This is a unique observation,
as one can study in general only the dynamics of critical 
states, the end state, but not critical states in the making, 
viz while they are forming \cite{markovic2014power}.

The formation of log-normal and powerlaw distributions can
be interpreted within an information-theoretical approach
to human activities, that we present. Within this approach 
human activities are assumed to produce maximum entropy 
distributions, that is distributions for which the information 
content is maximal. The exponential distribution, which is 
entropy-maximal under the constraint of a given mean, becomes 
a powerlaw once the Weber-Fechner law is taken into account, 
namely that the brain discounts sensory stimuli, numbers and 
time logarithmically.
\cite{hecht1924visual,dehaene2003neural,howard2018memory}.
Considering next that people differ with respect to their
preferences, which includes having distinct expectations for
the mean of the distribution to be generated, one obtains a
log-normal distribution. Using this framework we propose that
the observed change from a log-normal lifetime distribution
to a powerlaw is due to a decoupling of the individual time
horizons from decision making. There is no need to plan a trip
to the next music store, to illustrate this statement, when 
an album can be bought on the spot, online, once somebody has
discovered a song of her or his liking.

Music charts come in two varieties. As weekly sales charts,
which is typically the case for albums, and as airplay charts,
for which the number of times a song is aired by radio
stations is counted. For airplay data, which are often 
included for single charts, the underlying generative process 
is the decision making of a restricted number of radio operators.
Sales statistics results in contrast from the collective
behavior of a potentially very large number of individuals.
It is hence not surprising that the statistics of airplay and
sales charts differ, as we find, on a fundamental level.
The tendency to self-organize observed for sales charts
does not manifest for airplay charts. In this study we 
concentrate on sales, viz album charts.

Within the last decade most algorithms used to determine
chart rankings have been updated with respect to the
inclusion of streaming and downloads. We find that
streaming tends to reduce both the number of albums
making it to a chart, the chart diversity, and the
inner mobility, that is the average weekly rank changes.

\begin{figure}[t]
\centerline{
\includegraphics[width=0.90\columnwidth]{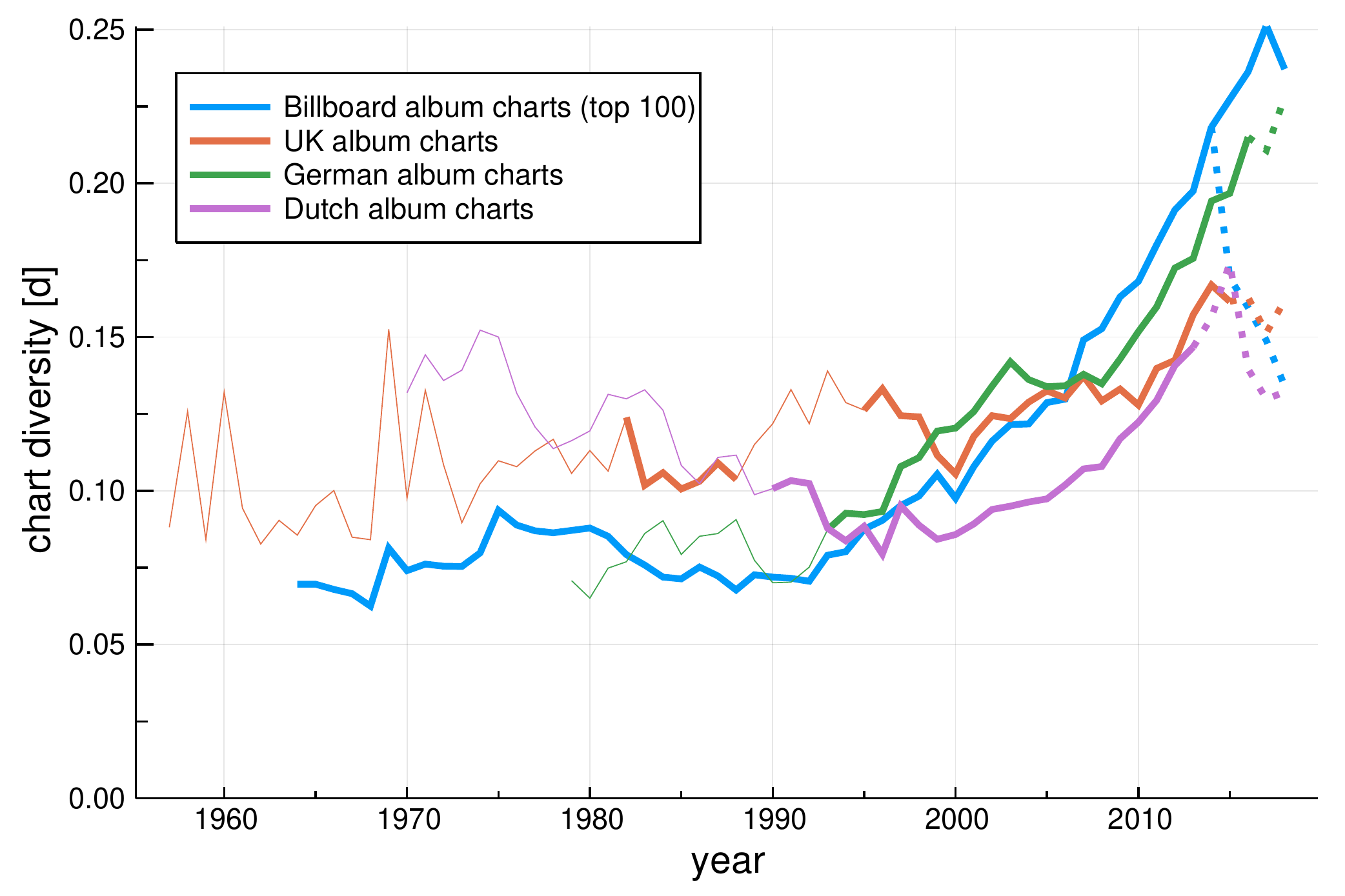}
           }
\caption{{\bf Chart diversity.} The evolution of the 
chart diversity, which is defined as the fraction 
$d\!=\!N_a\!/\!N_s$ of the number of distinct albums 
$N_a$ listed in a given year and the number $N_s$ of 
slots available. Lines are thin for periods for which less
than 100 chart positions are available, and dashed once
streaming was included. For a top 100 chart and 52 weeks
per year there are $N_s\!=\!100\!\cdot\!52$ slots. One observes
that the chart diversity has increased steeply for the US and the
German sales charts, in particular since 1990. For the
Billboard album charts the original sales-based rankings metric 
is available under a new name (full line), as of 2014/15, 
together with the updated version that is based on a multi-metric 
consumption rate (dashed line). The average number of weeks a song 
remains in the chart in a given year is of the order of 
$1\!/\!d$, compare Fig.~\ref{meanLifetime40}.
}
\label{chartDiversity}
\end{figure}

\section{Results\label{sect_results}}

The US billboard charts, the UK charts, the German
and the Dutch music charts were obtained
from public internet sources
\cite{billboardCharts,britishCharts,germanCharts,dutchCharts}.
An important parameter is the length of a chart,
which typically increased over the years. For
quantities that can be normalized with respect
to the number of entries available in a given year,
the entire timeline can be examined. For other
features, quantities that depend on absolute
and not on relative rankings, we restricted the analysis
to charts that list at least the top 40/100
albums, which has been the case since
1963/1967/1978/1979, and respectively since
1963/1982/1993/1990, for the Billboard, the UK, 
the German and the Dutch album charts. Somewhat
special are the US Billboard album charts, which
increased to $200$ in 1968. In the graphs we indicate, 
when suitable, whether top 100 or less chart ranks
where available.

The algorithms used for the compilation of the
individual charts have been adjusted over time,
mostly in minor ways. A major update occurred for the
Billboard charts in 2014/15, when the traditional
sales-based ranking was substituted by a ranking
based on a multi-metric consumption rate, which
includes weighted song streaming. This update,
which took effect end of 2014, affected the chart
statistics profoundly. Streaming data was included
respectively since 2014/2016/2017 for the Dutch,
the UK and the German charts.

The original Billboard album chart, the Billboard Top 200,
was retained after the 2014/15 metric update under a new name, 
as `Top Album Sales'. Data continuity is consequently achieved
when using, as we have done, the Top Album Sales charts from 2014/15 
on. Whenever possible we will show results obtained from both the 
Billboard Top 200 and the Top Album Sales charts, where the latter 
are compiled according to the unaltered sales-based ranking rules, 
albeit only for 100 ranks. For the UK, German and Dutch charts
only the version including streaming is accessible after the
respective metric changes.

\begin{figure}[t]
\centerline{
\includegraphics[width=0.90\columnwidth]{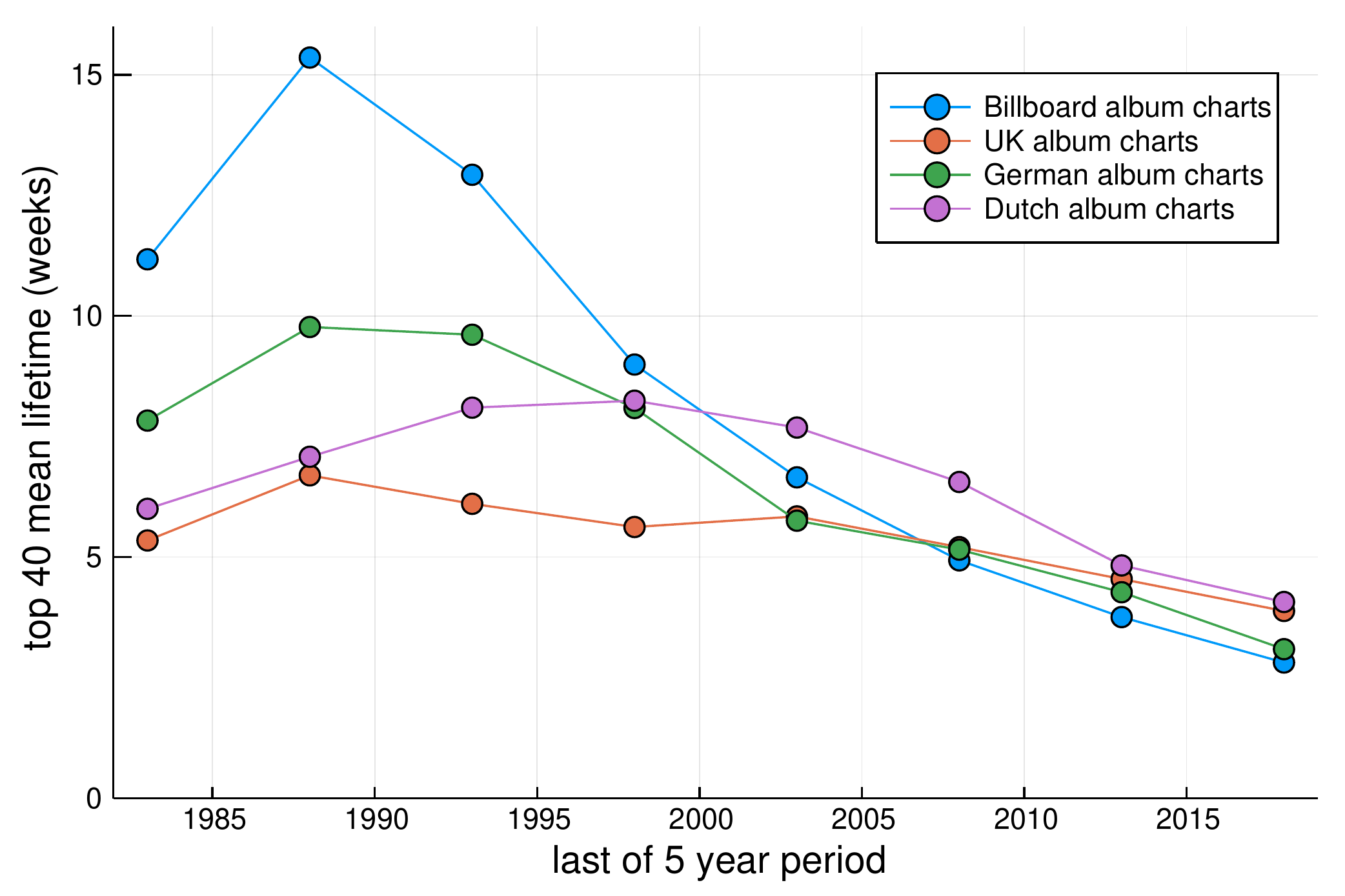}
           }
\caption{{\bf Album lifetime.} The top 40 mean lifetime,
namely the number of consecutive weeks an album is listed 
on the average among the top 40. The data has been pooled 
for trailing 5-year periods.
}
\label{meanLifetime40}
\end{figure}

\begin{figure}[t]
\centerline{
\includegraphics[width=1.0\columnwidth]{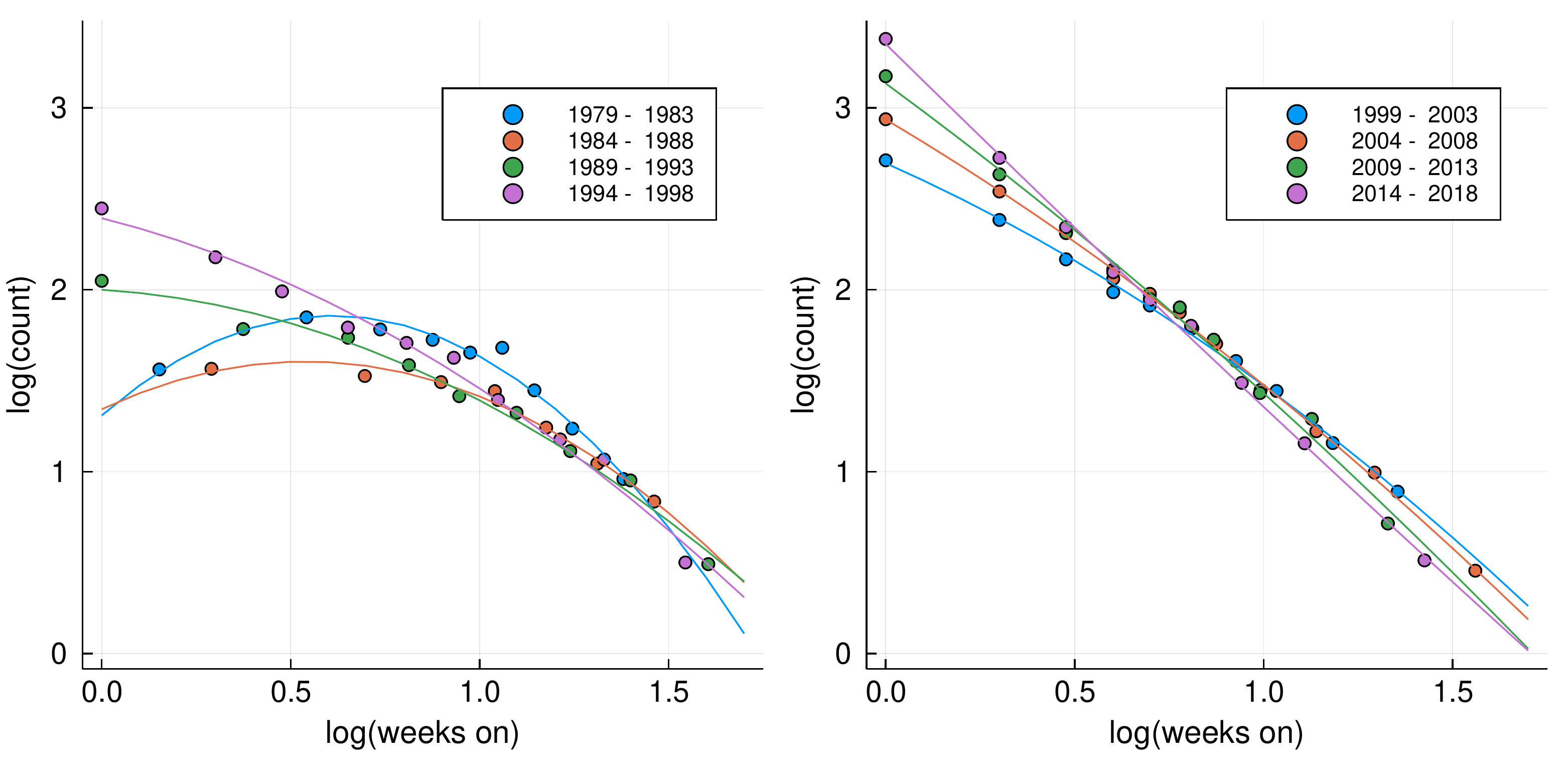}
           }
\caption{{\bf Lifetime distribution.} On a basis ten log-log plot,
the top 40 lifetime distribution, namely the distribution of
the number of weeks a given album is listed among the top 40 on the
Billboard chart. The data (circles) has been pooled for successive
5-year periods and fitted quadratically (lines), compare (\ref{log_normal}).
One observes that the lifetime distribution evolves over the years
from a log-normal distribution towards a power-law, which corresponds
respectively to a quadratic and a linear dependency in a log-log
representation.
}
\label{weeksOnBillAlbum}
\end{figure}

\subsection{Chart diversity - the negative effect of streaming}

For a first understanding of the data we examine the
evolution of the overall number of albums making it to
the charts in a given year. As a gauge for the
chart diversity we normalize the number of distinct
albums $N_a$ listed in a given year by the overall
number $N_s$ of available slots. A top 100 chart
could list, to give an example, a maximum of
$N_s\!=\!100\!\cdot\!52$ distinct albums per year. The
average number of weeks $\bar{w}$ an album is listed
in a given year is then just of the order of the
inverse chart diversity, $\bar{w}\!\approx\!1\!/\!d$. The
overall chart lifetime of albums will be discussed
further below.

In Fig.~\ref{chartDiversity} the evolution of the
chart diversity is presented on a year to year basis.
One notices that the US and German charts follow
qualitatively and quantitatively similar trends and
that the chart diversity increases rapidly since the
early 1990s. The average number of weeks $\bar w$ an
album was listed in a given year in these two countries 
decreased correspondingly from about $1/0.075\!\approx\!13$ 
in 1990 to $1/0.2=5$ in 2014/15. For this instance 
cultural processes accelerated by a factor more 
than two.

The Billboard charts data was split in 2014/15, when the 
traditional sales-based ranking was supplemented by
a multi-metric consumption rate that includes streaming.
For the latter the trend to become more diverse reversed.
Similar but less pronounced effects can be observed for 
the UK and the Dutch charts, but not for the German music 
charts. Note in this respect, that there are different routes,
as detailed in the Appendix, on how to include streaming 
and song downloads.

Over their entire histories, the diversity of the British 
and Dutch music charts does not show pronounced trends. 
However, as visible in Fig.~\ref{chartDiversity}, a measurable
increase in diversity is observed for the last two decades.
The underlying reason for the otherwise distinct behaviors
of the Dutch and UK charts with respect to the US and German
charts is at present not evident.

\subsection{Album lifetimes - a self-organized critical state in the making}

For all charts we have evaluated the number of weeks $n_w$
a given album remains within the listed range, the lifetime
of an album. The lifetime is an absolute number which is
not easily normalizable relative to the number of ranks
available. In order to be able to compare the four charts
over a comparatively long time span, we analyze only the
top 40 albums. This restriction allows us to go back till
1979, the year when the Dutch charts increased their length
to 40 ranks. For the Billboard charts we checked also the
long-term evolution for the top 100 albums, finding very
similar trends.

The mean album lifetime pooled over trailing 5-year 
periods is presented in Fig.~\ref{meanLifetime40}.
The top 40 lifetime is roughly inversely proportional
to the chart diversity shown in Fig.~\ref{chartDiversity},
which is however normalized to the chart length on a yearly
basis. For the Billboard charts the mean lifetime has seen
a reduction by more than a factor two over the last 25 year.

Our main focus is the lifetime distribution, which
is defined as the probability $P(n_W)$ for an
album to remain listed $n_w$ weeks. We find that $P(n_W)$
can be fitted quite accurately by a log-normal
distribution,
\begin{equation}
P(n_w) \sim \mathrm{e}^{-a\ln(n_w)-b\ln^2(n_w)}\,,
\label{log_normal}
\end{equation}
as shown as a log-log plot in Fig.~\ref{weeksOnBillAlbum}
for the US Billboard charts. One observes that the
lifetime distribution evolves steadily over the
course of roughly 4 decades, from a quadratic to a
linear dependency in a log-log representation. The
lifetime distribution reduces to a powerlaw
$P(n_w) \sim 1/(n_w)^a$ in the limit $b\to0$, an indication
of a critical state \cite{markovic2014power}.

\begin{figure}[t]
\centerline{
\includegraphics[width=1.00\columnwidth]{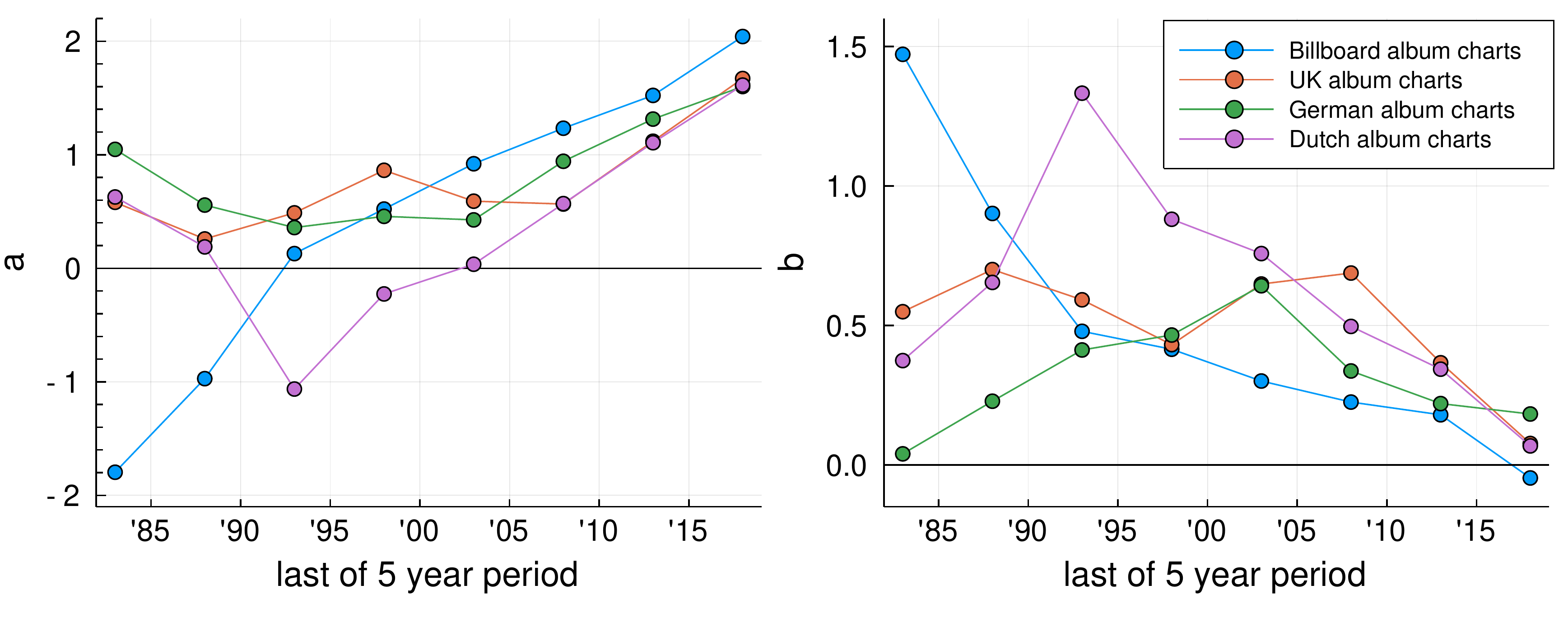}
           }
\caption{{\bf Coefficients of the lifetime distribution fits.}
The time evolution of the fit parameters $a$ and $b$ for the 
top 40 album lifetimes (circles), which correspond to the 
number of weeks $n_w$ an album is listed among the top 40.
The lifetime distribution has been fitted by
$\exp(-a\ln(n_w)\!-\!b\ln^2(n_w))$,
as illustrated in Fig.~\ref{weeksOnBillAlbum} for the
Billboard charts. This functional form corresponds to a
log-normal distribution for $b\!>\!0$ and to a powerlaw for
$b\!\to\!0$. The data is for trailing 5-year periods. Over
time, the lifetime distribution become more powerlaw-like,
with the exponent $a$ approaching the marginal value 
$a\!\to\!2$. For the German music charts the trend is 
less clear. The lines are guides to the eye.
}
\label{weeksOnBillAlbum_a_b}
\end{figure}

In Fig.~\ref{weeksOnBillAlbum_a_b} the evolution of
the fit parameters $a$ and $b$ entering (\ref{log_normal}) 
are shown for all charts investigated. One observes 
that $b$ tends to become small, in particular for
the last 30 years, with the exponent $a$ approaching 
$2$. We note that an exponent of 2 is marginal, as 
the mean of $P(n_w)$ diverges formally for $a<2$ and 
$b=0$. Nearly marginal exponents are however not 
uncommon \cite{markovic2014power}, with a well-known example
being the in-degree of domains in the world wide web
\cite{gros2012neuropsychological}.

The occurrence of powerlaws in evolving systems indicates 
the emergence of a self-organized critical state 
\cite{bak1987self}. In general it is to be expected 
that social systems, like the well studied network of 
scientific collaborations \cite{barabasi2002evolution},
are characterized by evolving parameters. The data
for the lifetime distribution presented in
Figs.~\ref{weeksOnBillAlbum} and
\ref{weeksOnBillAlbum_a_b} is however unique, in the
sense that it allows to examine not only the
final state, but the entire self-organizing process.
A derivation of (\ref{log_normal}) based on an
information-theoretical analysis will be
presented in Sect.~\ref{sect_IT_HA}.

\subsection{Number one albums -- the start determines the fate}

\begin{figure}[t]
\centerline{
\includegraphics[width=1.00\columnwidth]{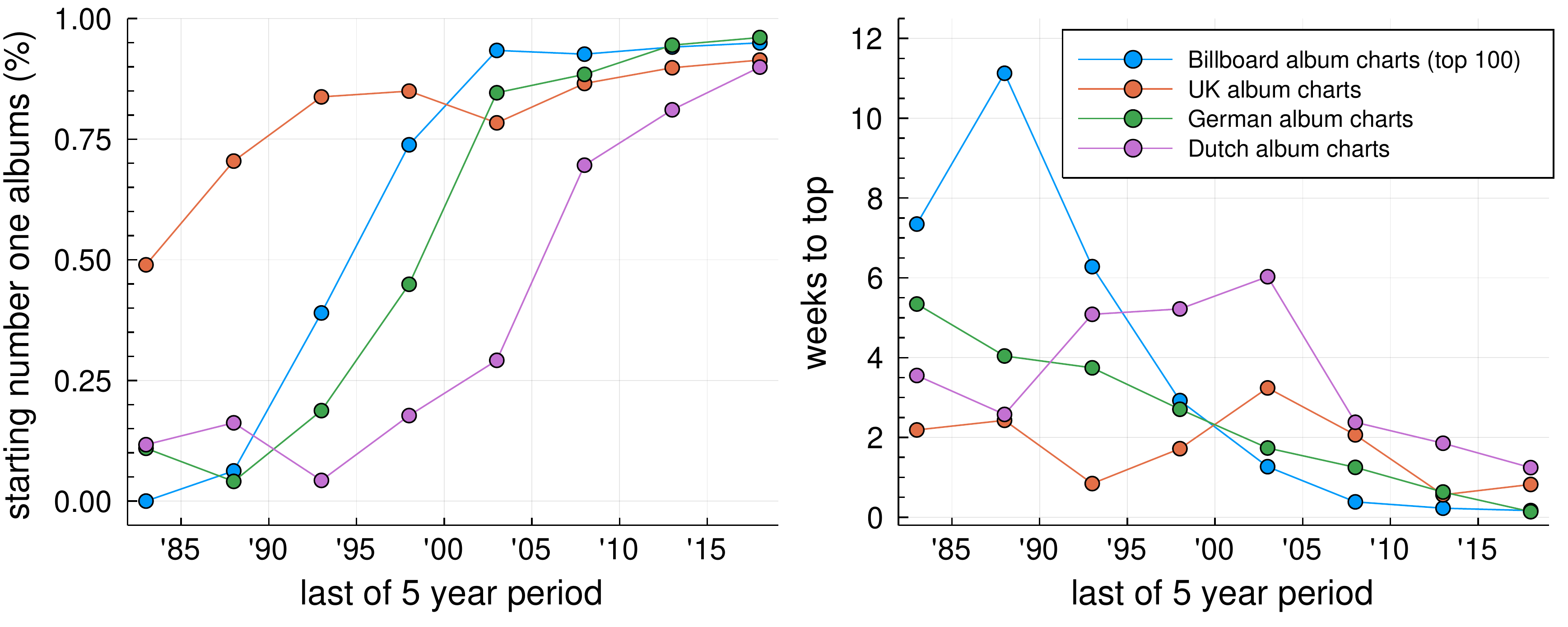}
           }
\caption{{\bf Number one albums.}
{\em Left:} The probability $P_{\mathrm{one}}$ that
a number one album started as such. The fraction of
albums managing to reach the top when starting form
a lower entry position is $1-P_{\mathrm{one}}$.
{\em Right:} The average number of weeks number one
albums did need to reach the top. Starting at the top
corresponds to zero weeks.
}
\label{numberOneAlbums}
\end{figure}

Commercial performance in terms of weekly sales varies
vastly between albums. Of key importance is in this
regard the first rank an album attains, the entry
position. In Fig.~\ref{numberOneAlbums} we present the
probability that a number one album started as such.
In the past essentially no album started on the top and
albums succeeding to reach the top could take a month or
more to do so. Today the situation is reversed, with the reversal
being nearly complete for the US, the German and the Dutch
charts, and somewhat reduced in magnitude for the UK charts.
A time lag of about a decade is furthermore observable
between the Billboard and the Dutch charts.

The rising predominance of number one entries is reflected
in the number of weeks an album needs on the average to
climb to the top, the climbing time. The zero for the
data shown in Fig.~\ref{numberOneAlbums} is set to the top,
which implies that the climbing time for albums entering at 
the top is zero.
It is quite remarkable that the average climbing time has
seen, modulo fluctuations, a dramatic decrease both for
the Billboard and the German music charts. This observation
holds to a certain extend also for the Dutch charts, but
not for the UK charts, which changed less over the last
three decades. This more conservative evolution of the
UK charts is consistent with the results for the chart
diversity shown in Fig.~\ref{chartDiversity}. Overall
we believe that the data presented in Fig.~\ref{numberOneAlbums}
provides convincing evidence that the market penetration of
new albums is now very fast, taking on the average a week
or less, depending on the country. Three decades ago, the
same process took about 2-3 weeks in the UK and more than
the double in the US.

In Table \ref{table_number_one_entries} we present for
the Billboard album charts a compendium of statistical
data that describes the dynamics of number one albums.
For the order of the average first and second week ranks
one observers a reversal in in the ordering. Before the
mid-90s albums climbed, afterwards the rank could only 
decay. Also evident is a substantial shortening of the
number of weeks at the top, which was defined here as the
number of weeks from the first to the last time an album 
was listed with a rank of one, including hence interruptions,
which are of the order of 10-20\%. As of today, albums are
given little time to stay at the top, as number of albums 
attaining the top rank in a given year has increased so 
strongly, by a factor 3-4 since the 80s, that the number
of yearly number one albums starts to approach the 
maximum of 52, compare Table \ref{table_number_one_entries}.

\subsection{Entry and exit positions - ongoing symmetrization}

\begin{figure}[t]
\centerline{
\includegraphics[width=1.0\columnwidth]{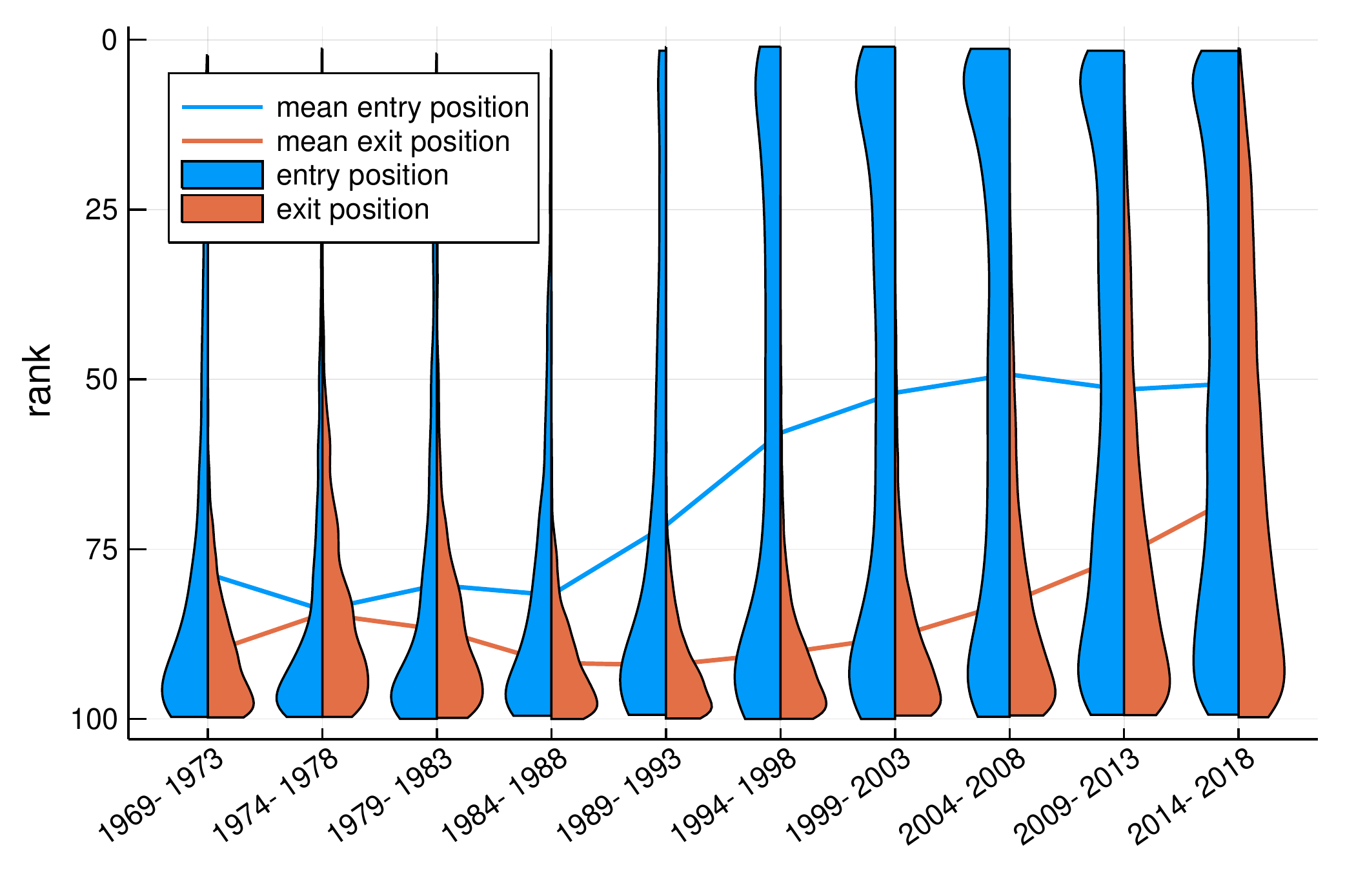}
           }
\caption{{\bf Entry \& Exit distributions.} For the top 100
Billboard album charts the distribution of entry (blue) and
exit ranks (orange), average over five years periods. The
width of the violin-charts measures the respective
probabilities. Also included are the mean entry and exit
positions (lines). Note the dramatic increase in top-ranked
entries in the 90s. The corresponding raise in exit ranks
has been more sequential. Checking for different chart lengths,
top 40 and top 200, we found a qualitative similar behavior that
is rescaled according to the chart length considered.
}
\label{entryExitDistributionBill}
\end{figure}

How likely is it, that a given album manages to climbs at
all, once in the charts? The evolution of the mean entry
ranks is shown in Fig.~\ref{entryExitDistributionBill}
for the Billboard charts, together with the mean exit positions.
For the full entry and exit distributions, also included
in Fig.~\ref{entryExitDistributionBill}, Gaussian-broaden
violin graphs have been generated, with the horizontal width
being proportional to the probability of finding an entry/exit
position within the respective five-year period. The data for the
UK, German and Dutch is similar, but in part less pronounced.

The first-listing ranks are due to external effects, such
as the quality of the album and the size and the penetration
speed of publicity campaigns. The distance of the average
exit ranks to the bottom, located at one hundred in our case,
is on the other side determined by the size of the average
inner mobility, for which we will provide a specific
definition in the next section. Here we point out that the
exit ranks are in general rising, which means that
the inner mobility is accelerating.

The full entry rank distribution presented in
Fig.~\ref{entryExitDistributionBill} is consistent
with the data for number one albums shown in
Fig.~\ref{numberOneAlbums}, in the sense that
the probability of higher entry positions
has been continuously increasing since the
early 90s. A remarkable and somewhat astonishing
result is the symmetry the entry distribution
exhibits nowadays with respect to 50, the half-way
rank between the bottom and the top. We checked that
this observation holds also for the entry distribution
of top 40 and top 200 albums.

\begin{table}[b]
\caption{{\bf Rank statistics of number one albums.} 
\label{table_number_one_entries}
Rank statistics for albums making it to the top. Data 
for the Billboard album charts, as averaged over 
trailing five-year periods (given is respectively the 
final year). Shown are the means for the entry rank (\#1 first), 
for the second week position (\#1 second) and for 
the exit rank (\#1 exit). Also given is are the number
of weeks to climb to the top (\#1 climb), the time at the
top (\#1 top), including interruptions, the number of 
weeks from top to exit (\#1 exit) and the average
number of number one albums per year (\#1 albums).
For comparison for all albums the mean for the 
entry and exits ranks (all first/all exit) are given,
together with the average album lifetime (all lifetime).
It is indicated whether the data is for the chart rank
(R) or the number of weeks (W). Note the order reversal 
of first and second-week ranks of number one albums between 
1998 and 2005. Compare Figs.~\ref{numberOneAlbums} and
\ref{entryExitDistributionBill}.
}
\centerline{\begin{tabular}{lcrrrrrrrrrr}
\hline
            &   & 1973 & 1978 & 1983 & 1988 & 1993 & 1998 & 2003 & 2008 & 2013 & 2018\\
\hline\hline
\#1 first   & R & 43.7 & 45.4 & 34.8 & 43.8 & 21.5 &  7.5 &  3.4 &  1.6 &  1.4 &  1.8\\
\#1 second  & R & 17.6 & 22.1 & 20.2 & 26.3 & 10.1 &  6.3 &  5.0 &  4.4 &  5.0 &  8.1\\
\#1 exit    & R & 91.7 & 86.3 & 89.5 & 93.6 & 92.2 & 93.6 & 91.9 & 89.9 & 85.1 & 74.9\\
\hline
\#1 climb   & W &  5.4 &  7.1 &  7.2 & 11.1 &  6.3 &  2.9 &  1.3 &  0.4 &  0.2 &  0.2\\
\#1 top     & W &  4.4 &  4.0 &  5.7 &  5.4 &  4.4 &  2.4 &  2.2 &  1.5 &  1.5 &  1.3\\
\#1 exit    & W & 34.6 & 24.5 & 30.5 & 40.9 & 43.7 & 37.1 & 33.0 & 29.8 & 28.5 & 23.5\\
\hline
\#1 albums  &   & 12.4 & 14.0 & 10.0 & 10.8 & 13.2 & 22.6 & 25.6 & 36.0 & 35.0 & 41.0\\
\hline
all first   & R & 78.6 & 83.9 & 80.1 & 81.7 & 71.5 & 57.9 & 52.0 & 49.3 & 51.5 & 50.6\\
all exit    & R & 90.2 & 84.9 & 86.4 & 91.7 & 92.2 & 90.3 & 88.1 & 83.0 & 76.3 & 67.6\\
\hline
all lifetime& W & 16.4 & 12.9 & 14.0 & 16.9 & 14.8 & 11.7 &  9.2 &  7.0 &  4.9 &  3.5\\
\end{tabular}}

\end{table}

\subsection{Inner mobility - accelerating rank decay}
\label{section_inner_mobility}

Once an album makes it to a chart, it may go up and down
on a weekly basis. We define the relative inner mobility
$M_I$ as
\begin{equation}
M_I = \left\langle \frac{R(t-1)-R(t)}{R}
\right\rangle,
\qquad\quad
R=\mathrm{max}(R(t),R(t-1))\,,
\label{M_I}
\end{equation}
where $R(t)$ is the rank a given album has in week $t$,
and where $\langle\cdot\rangle$ denotes the average over
all albums within a certain period. Entry and exit
weeks are not included. The max-function in (\ref{M_I})
ensures that $|M_I|<1$. For the sign we have two
cases.
\begin{itemize}
\item Climbing: $R(t)<R(t-1)$. The contribution to (\ref{M_I})
      is $(R(t-1)-R(t))/R(t-1)$, which is positive.
\item Descending: $R(t)>R(t-1)$. The respective term
      is $(R(t-1)-R(t))/R(t)$, which is negative.
\end{itemize}
Instead of $M_I$ one can study the absolute inner mobility
$R(t-1)-R(t)$, which would however weight an increase from 90
to 80 equal to an advancement from rank 11 to rank 1.

\begin{figure}[t]
\centerline{
\includegraphics[width=1.00\columnwidth]{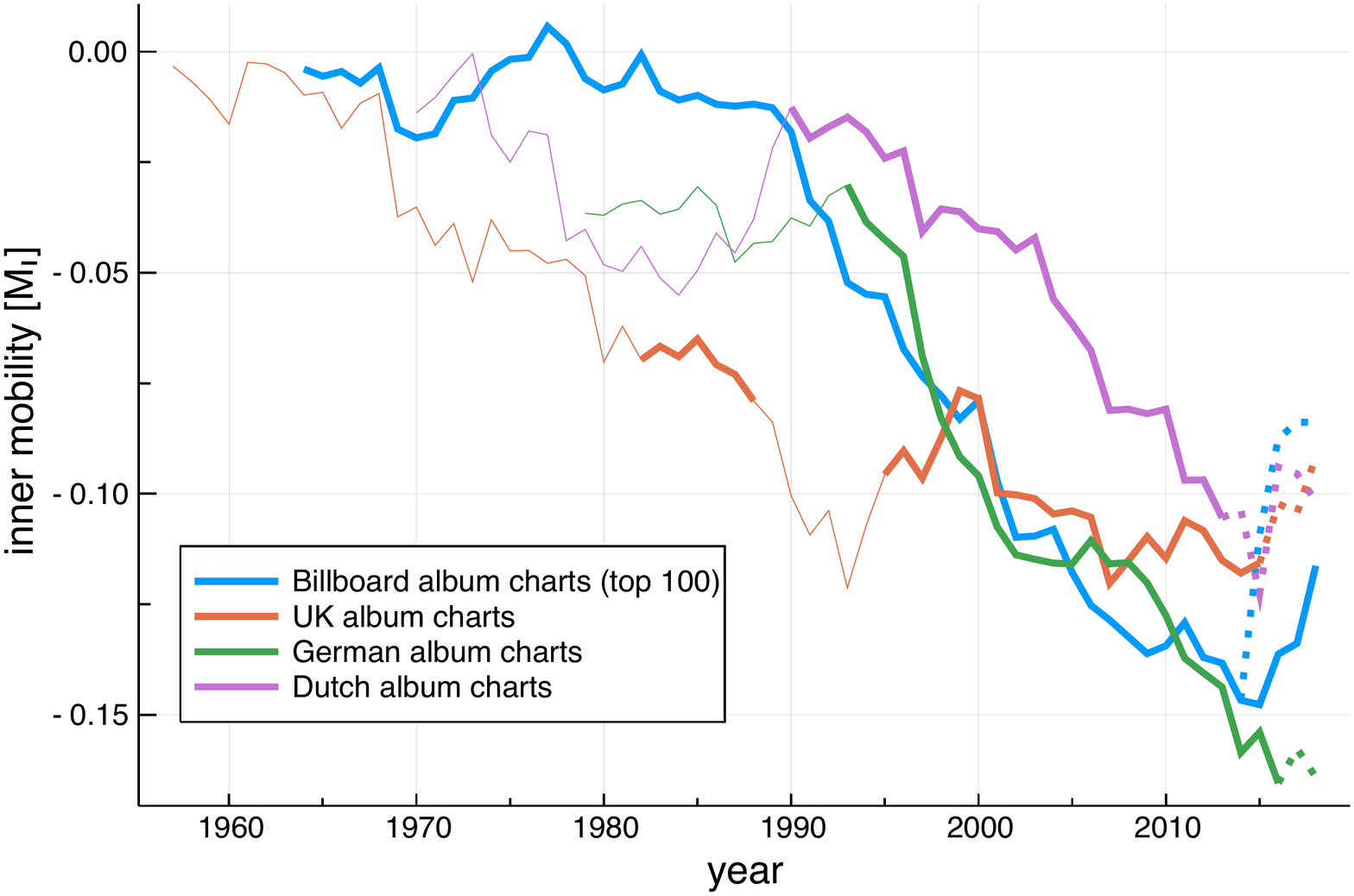}
           }
\caption{{\bf Inner Mobility.} On a year by year basis the
relative inner mobility $M_I$
of albums in the respective music charts, as defined by (\ref{M_I}).
Lines are thin for periods with less than 100 chart positions and
dashed once streaming was included. Shown are the average weekly
rank differences, in percentage. An $M_I$ of $-0.1$ corresponds
to a rank decay rate of 10\% per week, such as a decline from rank
9 to rank 10, or from 90 to 100. Entrance and exit weeks are not 
counted. For the Billboard album charts a version retaining the
original sales-based ranking metric remained available after the 
20014/15 update, compare Fig.~\ref{chartDiversity}.
}
\label{innerMobility}
\end{figure}

The data for the inner mobility presented in
Fig.~\ref{innerMobility} shows that albums mostly
lose rank on the average, namely that $M_I<0$. It is
also evident that the weekly rank loss tends to increase
in size over time before streaming was included.
Modulo substantial fluctuation, this is the case for all
four charts investigated, with the German and the US
charts trailing each other surprisingly close. Similar
downward trends are also observable for the Dutch and
the UK charts, which was not the case for the chart
diversity shown in Fig.~\ref{chartDiversity}. The
timeline for the Dutch music chart can be interpreted
by a time lag of roughly a decade. For the UK
charts $M_I$ decreased in contrast earlier,
already during the 70s and 80s.

With $M_I$ being equivalent to a weekly decay rate,
we can define the decay time $T_I\approx1/M_I$, which
measures the time scales of the inner dynamics. Between
1990 and 2010, $T_I$ increased by about a factor of
three for the the Billboard, the German and the
Dutch charts, with a similar acceleration happening
for the UK charts 20 years earlier. In terms of the
inner dynamics, all four charts indicate that cultural
time has been accelerating, albeit not necessarily
at the same time and at the same pace.

The observation that albums move down on the average
implies that the average exit position is below the
mean rank of first listings, which is consistent with
the average entry and exit positions presented in
Fig.~\ref{entryExitDistributionBill}.
Recently a class of model describing unidirectional growth
processes that are terminated by a hard reset has been studied
\cite{biro2018unidirectional}. It is an interesting question
whether the inner mobility, as presented in Fig.~\ref{innerMobility},
could be described by an analogous but inverse dynamics.

\subsection{Billboard single charts - why airplay statistics differ}

The Billboard single charts are based on a mix
of sales data, jukebox playing and airplay, where
the latter counts the number of times a song is
aired by radio stations. The relative contributions
have been adapted over the years, with a major change
occurring in 2013, when streaming was included. We find
that the Billboard album and single charts differ 
substantially with respect to their statistical
properties, presumable because the single charts include
airplay, whereas the album charts do not. Album sales
are the result of a large number of individual decisions,
whether to buy or not, which reflects an extended range of
individual preferences. It is on the other side up to a
relatively small group of radio operators to select the
mix of songs that is likely to induce the targeted audience
of the radio station to remain tuned in.

\begin{figure}[t]
\centerline{
\includegraphics[width=1.0\columnwidth]{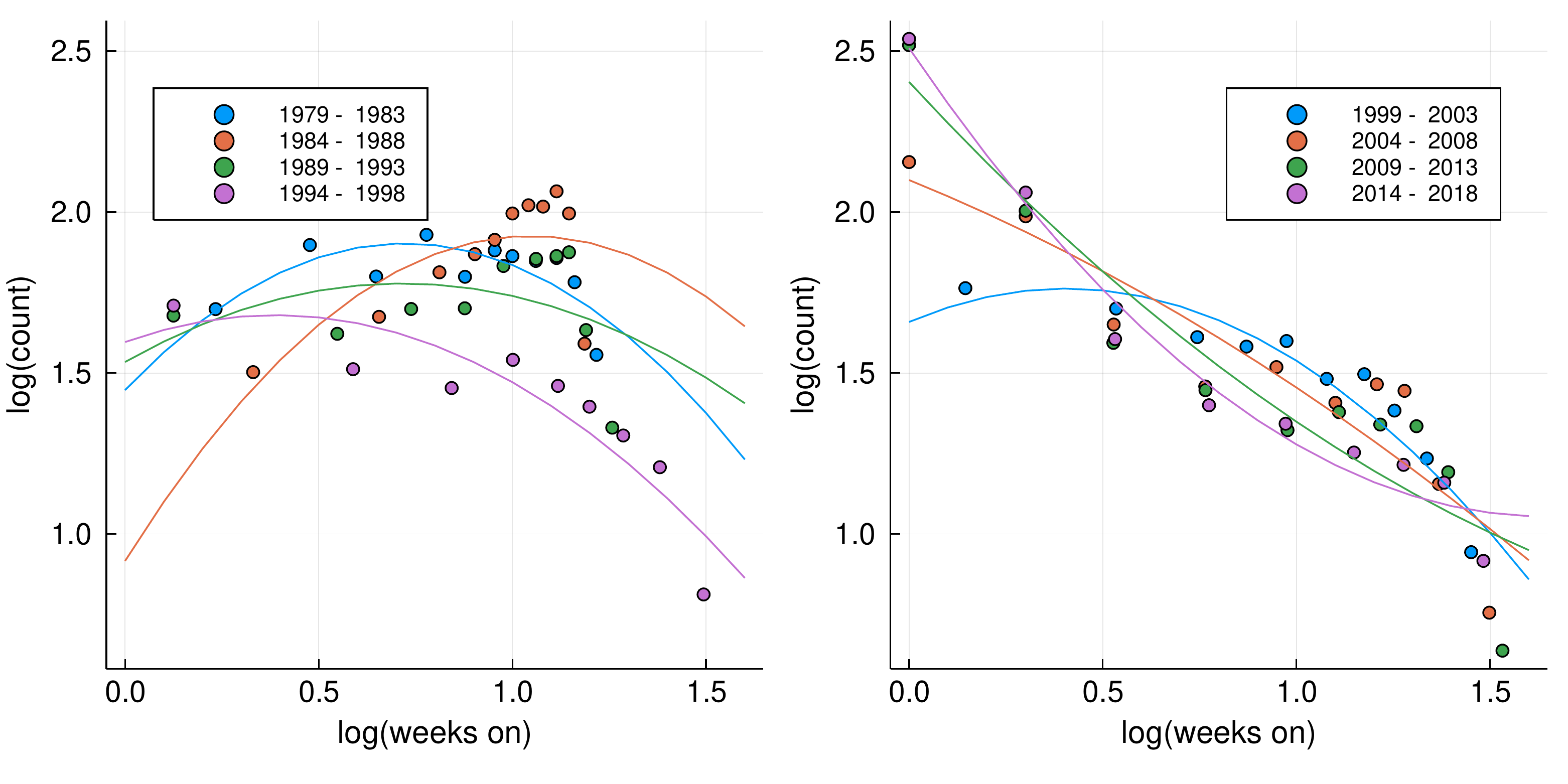}
           }
\caption{{\bf Single lifetimes.} On a basis ten log-log plot, the 
top 40 lifetime distribution for the Billboard single charts, which
are based in part on airplay data. Included are, as for Fig.~\ref{weeksOnBillAlbum},
least-square quadratic fits, which correspond to log-normal
and powerlaw distributions (\ref{log_normal}). It is evident that
the single lifetime distribution can not be approximated faithfully
by log-normal distributions or powerlaws.
}
\label{weeksOnBillSingle}
\end{figure}

One can see that the lifetime distribution of singles 
and albums are distinct when comparing
Figs.~\ref{weeksOnBillAlbum} and \ref{weeksOnBillSingle}.
The log-log plot of the top 40 single lifetimes
presented in Figs.~\ref{weeksOnBillSingle} shows that
quadratic fits are very poor, which implies that
single lifetimes cannot be described by (\ref{log_normal})
and that there is no evolution from a log-normal 
distribution to a powerlaw. A certain tendency for
the data to become more straight is however present,
possibly due to a crossover effect. Radio program
directors will be aware, in general, of the 
commercial success of the respective albums. Songs
from top ranked albums can hence be expected to enjoy 
a higher chance to be aired.

\subsection{International statistical convergence}

The picture emerging from our statistical analysis,
like the inner mobility, see Fig.~\ref{innerMobility},
is that the charts of two countries, Germany and the US,
show very similar trends. This would be trivially the case
if most of the songs making to the charts in Germany and
in the US would be the identical. A previous comparative
study of American, Dutch, French, and German popular
music charts found however no evidence for an ongoing
internationalization of popular music \cite{achterberg2011cultural}.
With regard to this question, which is not at the heart
of our present investigation, we note that a comparison of the
all-time most successful albums yields, as listed in
Table~\ref{table_top5} in the \nameref{section_appendix},
a similar result. Among the all-time top five German
albums two feature, to give an example, German `Schlager',
by Helene Fischer, with another one being a compilation
of German action songs for the Kindergarten,
`Die 30 besten Spiel...'. There is on the other side more
overlap between the all-time top 5 of the UK and the US album
charts, even though these two charts differ to a certain extend 
with respect to their overall statistical properties. A more
thorough investigation of this subject is left to future
studies.

\section{Information theory of human activities\label{sect_IT_HA}}

A large-scale study of the statistics of data
files publicly available on the internet showed
that the size distribution of formats having
a time dimension, like videos and audio files,
differs from static formats, such as jpeg and gif
images \cite{gros2012neuropsychological}. The
difference is that videos and audio files are
log-normal distributed, with the file-size
distribution of images following a powerlaw.
As a tentative explanation it was suggested that
the time domain corresponds for data files to
a second dimension, in addition to resolution,
and that the statistical distributions resulting
from human activities may be analyzed in many
instances from an information-theoretical perspective
\cite{gros2012neuropsychological}. Here we suggest
that the results presented in Figs.~\ref{weeksOnBillAlbum}
and \ref{weeksOnBillAlbum_a_b}, namely that the
statistics of album lifetimes evolved from
a log-normal distribution to a powerlaw, may
be analyzed along an analogous line of arguments.

Our underlying hypothesis is that there is a
feedback loop between the activities carried out
by a large number of individuals and the statistical
ensembles produced by these activities. For the case
of music charts this presumption implies that there is a
feedback between the lifetime distribution, which
results from people buying music, and the
individual decision to acquire a certain
album. E.g., the decision to go for an album
maybe influenced by the number of weeks the
album is already on the chart, and hence playing
on the radio. If this hypothesis holds, it is
reasonable to assume that the resulting
distribution should maximize information
in terms of Shannon's information entropy \cite{gros2015complex}.

\subsection{Human logarithmic discounting generates powerlaws}

The neurophysiological processes that give rise
to the ability of the human brain to process and
record information determine a subjective value
one attributes to an information source
\cite{hecht1924visual,buzsaki2014log,dehaene2003neural}.
This relation is known as the Weber-Fechner law. It states that the
neural representations of sensory stimuli \cite{hecht1924visual},
numbers \cite{nieder2003coding,dehaene2003neural,nieder2002representation},
and time \cite{takahashi2005loss,howard2018memory}, scale
logarithmically respectively with the intensity of the bare
stimulus, the number of objects and the length of a time span.

The Weber-Fechner law determines which type of distribution,
say of perceived stimuli $s$, is perceived to carry maximal
information. Consider that the neural working regime prefers
a certain mean for the perceived stimulus intensity, $\bar{s}$. 
The probability distribution $p(s)$ of perceived stimuli $s$
maximizing entropy is then an exponential,
$p(s)\sim\exp(-s/\bar{s})$. The Weber-Fechner law implies
that the external, the physical measurable stimulus $S$,
is logarithmically discounted, namely that $s=s_0\ln(S)$,
where $s_0$ is a characteristic scale. Using
$p(s)ds=p(S)dS$ one then finds
\begin{equation}
p(S) = p(s)\frac{ds}{dS},\qquad
p(S) \sim \frac{\mathrm{e}^{-s_0\ln(S)/\bar{s}}}{S} 
\sim \frac{1}{S^{a}},
\qquad\quad a=\frac{s_0}{\bar{s}}+1
\label{ME_S}
\end{equation}
for the maximal entropy distribution when expressed as 
a function of the afferent stimulus $S$. Maximization 
of information entropy under a logarithmic cost function 
yields hence generically a powerlaw, as shown here for 
the case of a single relevant variable. This viewpoint is 
complementary
to dynamical approaches, such as the reinforcement
loop via preferential attachment, that is the
`the rich get richer' principle \cite{barabasi1999emergence}.
For the case of music charts one has in consequence
that the lifetime distribution with the maximal
information content is a powerlaw.

\subsection{Entropy maximization with variable mean}

A maximum entropy distribution $\exp(-a s)$
is obtained by maximizing the objective function
\begin{equation}
\Phi(p) = -\int ds\, p(s)\ln(p(s)) - a\int ds s\,p(s)\,
\label{ME_Phi_s}
\end{equation}
where $p(s)$ is the probability density of $s$.
The first contribution to $\Phi(p)$ is the
entropy and the second the weighted average 
$\bar{s}=\int ds\,s\,p(s)$. 
The Lagrange parameter $a$ corresponds therefore 
to the relative weight of the average, the 
constraint. When $a$ is large the constraint
dominates maximization of $\Phi$.

We now assume that individuals differ with respect
to how much importance they give to album lifetimes
$s$, which will hence be reflected by the weight
of the mean album lifetime $\bar{s}$. For this
we introduce a hidden variable $h$, such that
individuals dispose of varying Lagrange parameters
$a\to (a+\kappa h)$, where $\kappa$ is a coupling parameter.
The joint distribution is then
\begin{equation}
p(s,h) \sim \mathrm{e}^{-(a+\kappa h)s} p(h)\,,
\label{ME_coupling}
\end{equation}
where $p(h)$ is the distribution of $h$, viz the
distribution of individual preferences. If we are 
interested only in the marginal distribution $p(s)$, 
which is typically the case when the hidden
variable $h$ is not observable, as in our case,
we obtain
\begin{equation}
p(s) \propto \int dh\, \mathrm{e}^{-(a+\kappa h)s} p(h)
     \sim    \int dh\, \mathrm{e}^{-(a+\kappa h)s}
                       \mathrm{e}^{-(h-\bar{h})^2/(2\sigma_h^2)}\,,
\label{ME_marginal}
\end{equation}
when assuming that $h$ is normal distributed with mean $\bar{h}$
and standard deviation $\sigma_h$. We set $\bar{h}\to0$
and absorb the mean $\bar{h}$ into the Lagrange multiplier $a$,
which can be done without loss of generality. One finds that
a Gaussian $p(h)$ leads to a Gaussian marginal $p(s)$,
\begin{equation}
p(s) \sim \mathrm{e}^{-a s-b s^2},
\quad\quad
b =\frac{(\kappa\sigma_h)^2}{2},
\quad\quad
s\propto \ln(S)\,,
\label{ME_gaussian_lognormal}
\end{equation}
which turns into a log-normal distribution, see 
(\ref{log_normal}), once the Weber-Fechner log-discounting
$s\propto \ln(S)$ is taken into account.

This result, Eq.~(\ref{ME_gaussian_lognormal}), suggest
that distributions of observables generated by the activity
of a large number of individuals are log-normal when there
is a substantial variability $\sigma_h$ of the
perceived individual means $1/(a+\kappa h)$.
One may view (\ref{ME_gaussian_lognormal})
as an alternative interpretation of the well known
result that Gaussians are maximum entropy distributions
when both the first and the second moment, mean and variance,
are given \cite{gros2015complex}. A maximum entropy distribution
with an optimized mean and variance is hence equivalent to
a maximum entropy distribution for which the optimized mean
is variable.

The log-normal distribution (\ref{ME_gaussian_lognormal})
evolves into a powerlaw for $b\to0$, viz in two cases,
$\kappa\to0$ and $\sigma_h\to0$. The first case, $\kappa=0$,
implies that the hidden variable does not couple to the
observable in first place, having hence no effect. All
individuals are identical in the second case, $\sigma_h=0$.
We note that $\Phi(p)$ is a functional of $p(s)$, 
which implies $\Phi(p)$ acts as a generating
functional, akin to the role generating functionals 
take in the context of guided self-organization
\cite{prokopenko2009guided,gros2014generating},
such af for attractor relict networks \cite{linkerhand2013generating}
and Hebbian learning rules \cite{echeveste2014generating}.
\subsection{Time horizons are less important when time accelerates}

People differ substantially in behavioral relevant
traits, such as the perception of time \cite{bartholomew2015analysis}.
The observation that individual likings are
 caused in general by a multitude of factors
\cite{gravetter2016statistics} suggests, that 
the distribution of preferences can be approximated 
by a Gaussian and that (\ref{ME_marginal})
constitutes a faithful representation of a maximum
entropy distribution when individual expectations vary.

For the case of music charts, we concentrated
on the album lifetime as our primary observable.
Chart rankings and the lifetime are determined
by weekly buying decisions that depend both on
a range of external factors, such as prominent marketing
campaigns, but also on the performance of albums
on the chart. We postulate here that the hidden variable
entering the information-theoretical interpretation
via (\ref{ME_marginal}) is related to the individual
perception of time, the time horizon. A log-normal
distribution would then be observed when a substantial
coupling $\kappa$ to the individual time horizons
is present. In this case it would matter, for a
buying decision, how long the album in question
has already been listed, and hence aired by
radio stations and on the internet. A powerlaw
is recovered on the other side when the
coupling to the individual time horizons 
ceases to be relevant.

The presumption of a decreasing relevance of the
time domain implies that one-time effects suffice
increasingly to influence buying decisions. This scenario
is not unlikely, given that the rise of the internet opened
the possibility to buy music essentially on the spot,
e.g.\ directly after one has heard or discovered a song,
online or on the radio. There is no need to plan
for a trip to the next music store on a free afternoon,
an undertaking that easily led in the past to delays
of days and weeks, and with this to a coupling between
buying and the personal time horizon. We stress, however,
that the arguments for a progressive decoupling of
the time horizon are at present only circumstantial
and that we cannot rule out that other drivings may
cause the observed changes in
the chart statistics.

Our argument, that personal time horizons may have
seen a decoupling from buying decisions draws
support also from the increasing relevance of top
entry ranks, as shown in Fig.~\ref{numberOneAlbums}
for number one albums. Before 1990 only very few
albums succeeded to enter the charts as number
one, which implies that publicity and commercial
success needed time, several weeks at least.
The situation has changed since the advent of the internet,
which allows news about new releases to spread
very fast via social media channels. For most
people buying an album will not affect the monthly
budget substantially, which is hence a decision
that can be carried out without further evaluation
once taken. The influence of the time domain
on buying albums is hence reduced.

\section{Discussion: Political time scales and stability}

One may ask whether the acceleration of the
information flow observed here for the case 
of music charts may be the reverberation of 
an equivalent speedup of societal and political 
processes at large. In this regard we point 
out that democracies rely in general on a 
stable and continuous evolution of public 
opinion \cite{enns2013public} and that it has
been suggested that not only the content of the 
political discourse is what matters, but in particular 
also the speed at which public opinion changes
\cite{gros2017entrenched}. 

Cultural and political processes condition each 
other \cite{street2015politics}, which 
implies that the respective time scales couple
\cite{rosa2005speed} and that social acceleration
will induce, if present and ongoing, a growing mismatch 
between political time delays, which are entrenched in a 
representative democracy by the electoral cycle,
and the accelerating pace of opinion dynamics. This is
an observation with potentially far reaching consequences,
as it is known from dynamical systems theory that 
a mismatch of instantaneous and delayed feedback
induces instabilities \cite{gros2015complex}. The 
outlook is then, from a dynamical systems perspective,
that modern democracies become inevitable unstable once 
the time scale of public opinion formation is shorter than
the time delays characterizing the interaction between
electorate and political decision making \cite{gros2017entrenched}.
Whether democracy as such is already in crisis is a question 
of debate \cite{plattner2015democracy,schmitter2015crisis,huq2018lose}.


\section{Conclusion}

Book, music and other charts are compiled in order to satisfy
the unabated interest \cite{fraiberger2018quantifying,liu2018hot}
in the commercial and artistic success of music albums, as well
as in other products of popular and classical culture.
They provide a valuable source for long-term socio-artistic
studies as their fundamental ranking criterion, success, has
not changed over the last 50 years, albeit modulo technical
adjustments. Given the continuity of the ranking metric,
changes of the chart statistics occurring over the time 
span of several decades are therefore reflecting
long-term socio-cultural developments.

We find three major trends. Firstly, one observes a 
substantial increase in the overall number of albums 
making it to the chart on a yearly basis, the chart 
diversity. The number of number one albums increased 
even stronger, it is nowadays around 40 per year for 
the Billboard album charts. Secondly, the route to become 
a number one hit has changed dramatically. Instead of 
climbing arduously from a modest entry rank, number one 
hits start nowadays as such. Finally we observe that the 
statistics of album lifetimes has seen a conspicuous change, 
evolving over the course of several decades towards a critical 
state. To our knowledge this is the first instance that the 
self-organization process as such may be studied explicitly
\cite{markovic2014power,gros2015complex}.
Within a proposed information-theoretical approach to 
human activities, the resulting powerlaw distribution 
of album lifetimes is due to the growing irrelevance of 
individuality, in the sense that the time necessary 
to form an opinion on whether to acquire an album, and 
to buy it, is now very low. It does not matter if somebody
needs only a few minutes to decide to download an album, 
or as long as a few days, as both timescales are
below the chart frequency of one week.

From an additional angle one can interpret the acceleration 
of chart processes found in this study as a measurable
indication that the cultural and social exchange of 
information occurs nowadays at a substantially faster 
rate than it used to. While intuitive, this observation could 
imply that the pace of opinion formation may have accelerated 
likewise over the past five decades. This would be a worrisome 
result, as it has been reported that representative democracies 
need to deal with the growing mismatch between the time delays
inherent in political decision making and an ever faster
opinion dynamics \cite{gros2017entrenched}, or face an
uncertain future.









\begin{table}[b!]
\caption{The top 5 albums with the longest lifetimes (overall weeks on charts),
for the Billboard charts (taking into account either the top 200
or the 100 ranks), and for UK, German and Dutch album charts. Listed
is also the respective number of consecutive weeks and the achieved top rank,
respectively from 1964/1957/1979/1970 until 2018.
\label{table_top5}
}

\begin{center}
\begin{adjustbox}{angle=90}
\begin{tabular}{cccccc}
\hline
chart & artist & title & weeks (all) & weeks (con) & top rank\\
\hline
\hline
Billboard  (200) & Pink Floyd & The Dark Side Of The Moon & 939 & 593 & 1\\
                 & Bob Marley And The Wailers & Legend: The Best Of... & 552 & 260 & 5\\
                 & Journey & Journey's Greatest Hits & 542 & 156 & 10\\
                 & Metallica & Metallica & 513 & 281 & 1\\
                 & Guns N' Roses & Greatest Hits & 450 & 138 & 3\\
\hline
Billboard  (100) & Metallica & Metallica & 286 & 163 & 1\\
                 & Adele & 21 & 260 & 144 & 1\\
                 & Bob Marley And The Wailers & Legend: The Best Of... & 247 & 31 & 5\\
                 & Kendrick Lamar & Good kid, m.A.A.d city & 236 & 56 & 2\\
                 & Imagine Dragons & Night Visions & 232 & 112 & 2\\
\hline
UK  & ABBA & Gold - Greatest Hits & 865 & 125 & 1\\
    & Queen & Greatest Hits & 838 & 224 & 1\\
    & Bob Marley And The Wailers & Legend: The Best Of... & 801 & 159 & 1\\
    & Fleetwood Mac & Rumours & 762 & 95 & 1\\
    & Meat Loaf & Bat Out of Hell & 520 & 203 & 9\\
\hline
German  & Andrea Berg & Best Of & 352 & 269 & 18\\
        & Helene Fischer & Best Of Helene Fischer & 337 & 301 & 2\\
        & ABBA & Gold - Greatest Hits & 317 & 60 & 1\\
        & S.~S., K.~G. \& die Kita-Fr\"osche & Die 30 besten Spiel- und B.-lieder & 290 & 33 & 43\\
        & Helene Fischer & Farbenspiel & 245 & 213 & 1\\
\hline
Dutch  & Adele & 21 & 330 & 167 & 1\\
       & Andr\'e Hazes & De Hazes 100 & 303 & 196 & 2\\
       & Buena Vista Social Club & Buena Vista Social Club & 294 & 172 & 7\\
       & Andr\'e Hazes & Al 15 jaar gewoon Andr\'e & 257 & 58 & 3\\
       & Dire Straits & Brothers In Arms & 248 & 171 & 1\\
\hline
\end{tabular}
\end{adjustbox}
\end{center}
\end{table}
\section{Appendix\label{section_appendix}}

In Table~\ref{table_top5} the top 5 most successful albums
are listed, according to the overall lifetime, the number
of weeks on the respective charts. For the US Billboard charts
which have a length of 200, the album lifetimes have been evaluated
taking into account either 100 or 200 ranks. For the UK, the German
and the Dutch album charts the respective entire data has been
evaluated. Also given is the number of consecutive weeks and
the respective highest rank reached.

Complementing the major rule updates given
in Sect.~\ref{sect_results}, we present here
for completeness an extended history of the 
Billboard album charts \cite{billboard_article_09,billboard_article_14}.
\smallskip

{\bf 1963}: The Billboard album charts start with top 150.

{\bf 1967}: Extension first to 175, and then to 200.

{\bf 1991}: \parbox[t]{0.8\textwidth}{
            Data source changed from phone call sampling of record stores
            \newline to `Nielsen SoundScan'.
            }\newline

{\bf 2010}: \parbox[t]{0.8\textwidth}{
            Catalog albums (older than 18 month, rank below 100, no single)
            \newline
            are included. Previously they were dropped.
            }\newline

{\bf 2014}: \parbox[t]{0.8\textwidth}{
            Metrics changed from a sales-based ranking to one measuring 
            multi-metric consumption, which includes streaming. 
            Sales-based album chart takes the name `Top Album Sales'.
            }\newline


The weighting of streaming can be performed along several routes.
When introduced for the Billboard album charts, 10 song sales or 1500 
song streams from an album were treated as equivalent to one purchase 
of the album. This changed 2018, when 1250 premium audio streams, 
3750 ad-supported streams, or 3750 video streams were consider to
equal one album unit. For the UK charts only the 12 most streamed 
songs of an album contribute instead, and not all. The weekly 
revenue and not the number of downloads enter on the other side 
the German charts.
        

\vskip1pc
\ethics{Not applicable.}

\dataccess{The paper deals with an analysis of music charts. 
The source of the data is given explicitly, references
25,26,27,28. Readers can access the data source, as it 
is publicly available.}

\aucontribute{Data acquisition and primary analysis by L.Schneider,
              study concept and interpretation by C.Gros.}

\competing{Not applicable.}

\ack{CG thanks Nathan~Valent\'\i\ for discussions.}

\funding{Not applicable.}

\disclaimer{Not applicable.}


\vfill
\pagebreak

\end{document}